\documentclass[12pt]{article}
\usepackage{amssymb,epsf}
\def\lsim{\mathrel{\rlap {\raise.5ex\hbox{$ < $}}
{\lower.5ex\hbox{$\sim$}}}}
\def\gsim{\mathrel{\rlap {\raise.5ex\hbox{$ > $}}
{\lower.5ex\hbox{$\sim$}}}} 
\usepackage{epsf}
\def\sqr#1#2{{\vcenter{\vbox{\hrule height.#2pt

        \hbox{\vrule width.#2pt height#1pt \kern#1pt

           \vrule width.#2pt}

        \hrule height.#2pt}}}}

\def\lsim{{\displaystyle
{{\raise-8pt\hbox{$ <$}}
\atop{\raise5pt\hbox{$\sim$}}}}}
\def\gsim{{\displaystyle
{{\raise-8pt\hbox{$ >$}}
\atop{\raise5pt\hbox{$\sim$}}}}}
\def\slsim{{\displaystyle
{{\raise-8pt\hbox{$\scriptstyle <$}}
\atop{\raise5pt\hbox{$\scriptstyle \sim$}}}}}
\def\sgsim{{\displaystyle
{{\raise-8pt\hbox{$\scriptstyle  >$}}
\atop{\raise5pt\hbox{$\scriptstyle \sim$}}}}}
\newskip\humongous \humongous=0pt plus 1000pt minus 1000pt

\newcommand{\sumpf}[0]{\sum_{(H^{\rm f},G^{\rm f})}\! \! \! \!
{\raise
4pt
\hbox{$'$}}\,}

\newcommand{\sump}[0]{\sum_{(H,G)}\! \! {\raise 4pt \hbox{$'$}}\,}

\def\bs{\begin{subequations}}
\def\es{\end{subequations}}
\catcode`\@=11
\newcount\hour
\newcount\minute
\newtoks\amorpm
\hour=\time\divide\hour by60
\minute=\time{\multiply\hour by60 \global\advance\minute by-\hour}
\edef\standardtime{{\ifnum\hour<12 \global\amorpm={am}%
        \else\global\amorpm={pm}\advance\hour by-12 \fi
        \ifnum\hour=0 \hour=12 \fi
        \number\hour:\ifnum\minute<10 0\fi\number\minute\the\amorpm}}
\edef\militarytime{\number\hour:\ifnum\minute<10 0\fi\number\minute}
\def\draftlabel#1{{\@bsphack\if@filesw {\let\thepage\relax
   \xdef\@gtempa{\write\@auxout{\string
      \newlabel{#1}{{\@currentlabel}{\thepage}}}}}\@gtempa
   \if@nobreak \ifvmode\nobreak\fi\fi\fi\@esphack}
        \gdef\@eqnlabel{#1}}
\def\@eqnlabel{}
\def\@vacuum{}
\def\draftmarginnote#1{\marginpar{\raggedright\scriptsize\tt#1}}
\def\draft{\oddsidemargin -.2truein
        \def\@oddfoot{\sl preliminary draft \hfil
        \rm\thepage\hfil\sl\today\quad\militarytime}
        \let\@evenfoot\@oddfoot \overfullrule 3pt
        \let\label=\draftlabel
        \let\marginnote=\draftmarginnote
   \def\@eqnnum{(\theequation)\rlap{\kern\marginparsep\tt\@eqnlabel}%
\global\let\@eqnlabel\@vacuum}  }
\def\subequations{\refstepcounter{equation}%
  \edef\@savedequation{\the\c@equation}%
  \@stequation=\expandafter{\theequation}
  \edef\@savedtheequation{\the\@stequation}
  \edef\oldtheequation{\theequation}%
  \setcounter{equation}{0}%
  \def\theequation{\oldtheequation\alph{equation}}}

\def\endsubequations{\setcounter{equation}{\@savedequation}%
  \@stequation=\expandafter{\@savedtheequation}%
  \edef\theequation{\the\@stequation}\global\@ignoretrue
  \vspace*{-12pt} \\}
\def\bs{\begin{subequations}}
\def\es{\end{subequations}}
\relax
\def\thefootnote{\fnsymbol{footnote}}
\def\be{\begin{equation}}
\def\ee{\end{equation}}
\def\ba{\begin{eqnarray}}
\def\ea{\end{eqnarray}}

\def\ee{\end{equation}}
\def\bea{\begin{eqnarray}}
\def\eea{\end{eqnarray}}

\newcommand{\uarrw}[0]{\mathrel{
{\raise.5ex\vbox{\hrule width 1cm}\hskip-6pt\rightarrow}}}
\def\thebibliography#1{%
\vskip 0.5cm \centerline{\bf References}
\list{%
[\arabic{enumi}]}{\settowidth\labelwidth{[#1]}
\leftmargin\labelwidth
\advance\leftmargin\labelsep
\usecounter{enumi}}
\def\newblock{\hskip .11em plus .33em minus .07em}
\sloppy\clubpenalty4000\widowpenalty4000
\sfcode`\.=1000\relax}

\renewcommand{\theequation}{\arabic{section}.\arabic{equation}}

\renewcommand{\section}{\setcounter{equation}{0}\@startsection%
{section}{1}{0mm}{-\baselineskip}{0.5\baselineskip}%
{\normalfont\normalsize\bfseries}}

\renewcommand{\subsection}{\@startsection%
{subsection}{2}{0mm}{-\baselineskip}{0.5\baselineskip}%
{\normalfont\normalsize\slshape}}

\renewcommand{\subsubsection}{\@startsection%
{subsubsection}{2}{0mm}{-\baselineskip}{0.5\baselineskip}%
{\normalfont\normalsize\slshape}}

\begin{document}
%
\renewcommand{\theequation}{\arabic{section}.\arabic{equation}}
\begin{titlepage}
\begin{flushright}
HU-EP-02/41 \\
hep-ph/0209296 
\end{flushright}
\begin{centering}
\vspace{1.0in}
\boldmath
{ 
\bf \large On the Time Dependence of Fundamental Constants$^\dagger$
}
\\
\unboldmath
\vspace{1.7 cm}
{\bf Andrea Gregori}$^1$ \\
\medskip
\vspace{.4in}
{\it  Humboldt-Universit\"at, Institut f\"ur Physik}\\
{\it D-10115 Berlin, Germany}\\

\vspace{3.2cm}
{\bf Abstract}\\
\vspace{.2in}
\end{centering}
We discuss how the results of recent measurements on the  
spectra of quasars are predicted by the non-perturbative solution of 
String Theory proposed in Ref.~\cite{estring}.
\vspace{4cm}

\hrule width 6.7cm
\noindent
$^\dagger$Research supported by the EEC under the contract 
EU HPMF-CT-1999-00396.\\
$^1$e-mail: agregori@physik.hu-berlin.de

\end{titlepage}
\newpage
\setcounter{footnote}{0}
\renewcommand{\thefootnote}{\arabic{footnote}}

It is quite on debate today the question on whether the fine
structure constant $\alpha$ does change with time or not\footnote{The 
variation of this parameter is expected in several
cosmological scenarios (for instance Ref.~\cite{grojean}).}, 
and in particular what is the meaning \cite{moffat,moffat2,cf,aduff}
of certain recent experimental
observations made on the radiation coming from ancient regions
of our Universe, suggesting that this quantity could have been
lower at earlier times \cite{dfmw,metal,wetal}. 
This conclusion is essentially based on  
the observed decrease of the relativistic correction to the 
behavior of emission spectra, modifying the
spectra of several elements detected in quasars. 
According to \cite{dfmw,dfw}, these corrections implicitly depend on $\alpha$,
and their decrease can be referred to a decrease,
as we go back in time, of its value: ${\dot{\alpha} \over \alpha} < 0$. 
In this note we want to discuss
how, in the light of the proposal of Ref.~\cite{estring},
a decrease of the relativistic effects should be better viewed
as the consequence of the increase of masses, as we approach
earlier eras. We will see how taking into account this effect as predicted
in Ref.~\cite{estring} correctly reproduces
the order of magnitude of the observed deviation.

As discussed in Ref.~\cite{estring}, masses can be viewed as a quantum gravity
effect, basically related to the fact that the effective size of
the Universe relevant for our physics is bound by the horizon of observation.
Each different type of particle lives on an "orbifold" of
our space-time, and feels differently the size of it, or, if one wants,
the age of the Universe. This is the reason why we have a varied spectrum of 
masses, whose present values agree with those experimentally measured.
As discussed in Ref.~\cite{estring}, the time dependence
of masses is expected to be:
\be
m_{\rm i}(t) \sim t^{- n_{\rm i}} \, ,
\label{mt}
\ee
where ${\rm i}$ labels the specific kind of particle (electron, muon etc...),
and $n_{\rm i}$ is a number, such that at the present time, $t=t_0$:
\be
m_{\rm i}(t_0) = t_0^{- n_{\rm i}} \equiv m_{\rm i} \, ,
\ee
$m_{\rm i}$ being the present-day value of the mass \footnote{Expressions
(\ref{mt}) are given up to coefficients, that in our case are 
of order 1.}.   
As discussed, this behavior, ultimately connected with the
time-dependence of the cosmological constant, is by itself able to account for 
the accelerated expansion of the Universe, and leads to a correct prediction
for the red shift.  A consequence of the change of masses with time
is also the change with time of the fine structure constant, which
is by definition the value of the electric coupling at a certain mass scale,
e.g. the electron's mass scale.
However, the rate and sign of this change is not the one proposed in 
Ref.~\cite{metal,wetal,dfw}. Indeed, according to Ref.~\cite{estring},
the coupling is expected to vary logarithmically with time,
and to become stronger at earlier times.
The reason of the discrepancy with the observations of Refs.~\cite{wetal,dfw},
is that the latter assume masses to be constant. As a consequence,
they are not accounted in the evaluation of the time variation of
the  corrections to the spectra due to relativistic effects, eq.~(11)
and following of Ref.~\cite{dfw}. 
Indeed, the second line of the r.h.s. of eq.~(11)
of Ref.~\cite{dfw} should also include, besides a dependence on $\alpha$, 
also a dependence on $m_{\rm e}$,  the mass of the electron, 
or better, its effective mass   
once the center-of-mass value, the so called reduced mass,
has been subtracted \footnote{In the hydrogen
atom this is given by $m_{\rm e} = {m_{\rm e} m_{\rm p} \over m_{\rm e} 
+ m_{\rm p}}$. The possibility of referring to a change of this quantity 
the effect measured in Ref.~\cite{dfw} can be found also in 
Ref.~\cite{cf,cf2,fshu}, although not related to a specific prediction 
as in our case.}. Even without knowing the details of 
this contribution, it is not difficult to realize that this goes in the 
correct direction, because the second line of eq.~(11) should vanish
as the mass goes to infinity (or much before, to the Planck scale):
the heavier is the particle, the less relativistic it is.
Therefore, the mass must enter 
the corrections with some opposite power ${\rm q}$ with
respect to $\alpha$: ${ \alpha \over m^{\rm q}}$ . 
The limit $\alpha \to 0$ should then be
qualitatively equivalent to the limit $m_{\rm e} \to {\rm M}_{\rm P}\, ,  
\infty$. 
In Ref.~\cite{estring} we have seen
how precisely a change of masses is at the origin of the
expansion of the Universe.
According to that analysis, there is then a precise relation between
time derivative of masses and red shift.
What we want to observe here is the deviation from the red shift.
This deviation depends on the shape of specific atoms and molecules,
and its change in time is due to the fact that, 
as masses get lighter with time (or heavier
when going backwards), these systems become more (resp. less)
relativistic. 
Owing to the fact  that, at least at first order,
$\alpha$ and $m$ enter the relativistic corrections with powers
of different sign, we expect the relative rate of change with time
of such corrections, as due to the mass change, to be of the
same order of that computed, as
a function of $\alpha$, in Ref.~\cite{dfw}.
In the following, we discuss how the results of Ref.~\cite{estring}
enable to predict the correct order of magnitude for this effect.

Even without going into the details of a precise evaluation
of the (mass-dependent) relativistic effects,
we can get an idea of the order of magnitude
we have to expect for the ``non-universal''
contribution to the deviation of emission spectra, namely,
the part that cannot be trivially re-absorbed into a redefinition
of the red shift. In fact, what we are looking for
here is not its absolute value, for which we would need a 
very precise computation, but its relative rate of change with time.
In order to roughly evaluate it, we can proceed as follows.
As we said, any overall change of mass essentially goes into a change of 
red shift.  
This is simply the statement that a universal change of the mass
scale is in practice nothing more than a
change of ``unit of measure'' \cite{aduff}. However,
masses don't change all with the same rate:
our system consists essentially of two mass scales, the one of the proton
and the one of the electron. A change of the average mass, that of the 
center of mass, is directly related to the red shift.
Also a change of the electron's mass itself can be re-absorbed into
a contribution to the red shift.
However, through the reduced mass the emission spectra receive corrections 
depending on the ratio $m_{\rm e / p} \equiv {m_{\rm e} \over m_{\rm p}}$.
It is basically the change with time of this quantity what ultimately
is the responsible for the non-universal effect, not re-absorbable
into an overall mass scale shift. Although the precise
non-universal contribution depends on the specific type, and ``shape'',
of element under consideration, an evaluation of the relative
change due to the contribution of $m_{\rm e / p}$ should anyway give
us an upper bound on this non-universal deviation.
Were not for this contribution, all the rest, including the relativistic
effects, would give at first order
a relative rate of change re-absorbable into an overall mass scale change.
This accounts therefore for the main effect we have to expect.  
The relative rate of change of frequencies can then be separated as:
\be
{\dot{\nu} \over \nu}  \approx {\cal O} \left({\dot{\alpha} \over \alpha}
+ {\dot{m} \over m} \right) + 
{\cal O} \left( \dot{m}_{\rm e / p}  \right) \, .
\ee
The first term of the r.h.s. goes into a redefinition of the red shift,
while the second one, accounting for
the displacement from the red shift, is of the order of magnitude
of the effect we want to evaluate. Owing to the way we have derived it,
it appears in a ``universal'' form, apparently the same
for any kind of elements. The point is that what we have obtained with this
argument is only a bound on the deviation we can expect. 
It belongs also to hydrogen, where it has been identified
for what it is, a mass variation, in Ref.~\cite{ivanchik}.  
Inserting  now the mass dependence (\ref{mt}), we obtain:
\be
{\dot{\nu} \over \nu}  \approx
{\cal O} \left( t^{-1} \right) \, - \, {\rm q} (n_{\rm p} - n_{\rm e}) 
t^{(n_{\rm p} - n_{\rm e}) - 1} \, .  
\ee
The first term is the universal part,
that goes into the redefinition of the red shift.
The second term, accounting for the relative
time variation of ${m_{\rm e} \over m_{\rm p}}$,
gives precisely what in Ref.~\cite{dfw} has been
reported as ${\dot{\alpha} \over \alpha}$. Since $(n_{\rm p} - n_{\rm e}) 
\sim - {1 \over 20}$, at the present day this term is of the order of: 
\be
\sim  {{\rm q} \over 20} \times {1 \over 2000} \times {1 \over 1.5} \times
10^{-10} {\rm yr}^{-1} ~~
\approx {\rm q} \times {1 \over 6} \times 10^{-14} {\rm yr}^{-1} \, ,
\ee
of the order of magnitude of the value one obtains by extrapolating the result
of Ref.~\cite{dfw} to the present time \footnote{We recall that
${\rm q}$ is the relative power
of the first order dependence on the mass, with respect 
to the coupling $\alpha$, entering
in the relativistic corrections to the emission spectra.
To the purpose of the present discussion a precise knowledge of ${\rm q}$ 
is not essential. Naively, by considering the way these parameters
enter in the basic expression of the electric force,
$\ddot{r}  = \left( { e^2 \over m} \right) {1 \over r^2}$,
we expect anyway ${\rm q} = 1$.}.
We remark that the rate of variation of $m_{\rm e} / m_{\rm p}$
predicted in Ref.~\cite{estring}
is compatible with the one suggested in Ref.~\cite{ivanchik}. 
We leave for future work a detailed evaluation
of the corrections to the spectra. In order to make a more
refined test of the proposal of Ref.~\cite{estring}, besides
an improved precision in the  experimental information, we need also
more exact mass formulae.

\vspace{.3cm}

In Ref.~\cite{estring} 
we have seen how masses are essentially a quantum gravity
effect, as is the cosmological constant. We have also seen that
the variation with time of these two quantities are tightly related:
a variation of the cosmological constant reflects in fact into 
a variation of the inertia of matter \footnote{All these issues
are discussed in detail in Ref.~\cite{estring}.}. A consequence of
this is that the Universe evolves from an earlier
``quantum gravity'' phase, toward a future more ``relativistic'' phase.
The quantum gravity phase dominates the more and more as we go
backwards in time, and the horizon restricts toward the Planck scale.
In that phase, all masses lift toward  the Planck mass scale, the minimal
scale allowed by the Uncertainty Principle. In the opposite direction,
as time goes by, the horizon expands,
the quantum gravity effects decrease, as do masses, and
the kinetic energy increases, as does the rate of expansion
of the Universe.
This evolution is reflected in the observed change of the emission spectra,
where one can detect both the ``macroscopic'' effect, i.e. the expansion
of the Universe, and the ``microscopic'' one. According to Ref.~\cite{estring},
we refer both to a change of masses. 

\vspace{1.5cm}

\centerline{\bf Acknowledgments}

I thank C. Schiller for interesting discussions.

\vspace{1.5cm}



\providecommand{\href}[2]{#2}\begingroup\raggedright\endgroup

\end{document}